\documentclass[12pt,dvips]{article}

\usepackage{rotating}
\usepackage{epsfig}
\usepackage{amssymb}

\newcommand{\imag}{\Im {\rm m}}
\newcommand{\real}{\Re {\rm e}}

\def\lsim{\mathrel{\raise.3ex\hbox{$<$\kern-.75em\lower1ex\hbox{$\sim$}}}}
\def\gsim{\mathrel{\raise.3ex\hbox{$>$\kern-.75em\lower1ex\hbox{$\sim$}}}}
\newcommand{\s}{\\ \vspace*{-3.5mm}}

\begin{document}

\begin{titlepage}

\begin{flushright}
DESY 04-055 \\
IFT-04/12\\
KIAS-P04028\\
LU-ITP 2004/018\\[2mm]
{\today}
\end{flushright}

\vskip 1.5cm

\begin{center}
{\Large\bf H/A Higgs Mixing  \\[1mm]
           in CP-Noninvariant Supersymmetric Theories
}\\[10pt]
\vskip 1.5cm
{\large S.Y. Choi$^1$, J. Kalinowski$^2$, Y. Liao$^3$ and P.M. Zerwas$^4$}
\vskip 0.5cm
{\it
 $^1$ Department of Physics, Chonbuk National University, Chonju 561-756,
      Korea\\[1mm]
 $^2$ Institute of Theoretical Physics, Warsaw University, PL--00681 Warsaw,
      Poland\\[1mm]
 $^3$ Institut f\"ur Theoretische Physik, Universit\"at Leipzig,
     D-04109 Leipzig, Germany\\[1mm]
 $^4$ Deutsches Elektronen-Synchrotron DESY, D-22603 Hamburg, Germany
 }
\end{center}

\vskip 4.0cm

\begin{abstract}
\noindent For large masses, the two heavy neutral Higgs bosons are
nearly degenerate in many 2--Higgs doublet models, and particularly in
supersymmetric models. In such a scenario the mixing between the states can be
very large if the theory is CP-noninvariant. We analyze the formalism
describing this configuration, and we point to some interesting experimental
consequences.
\end{abstract}

\end{titlepage}


\renewcommand{\thefootnote}{\fnsymbol{footnote}}

\section{Introduction}
\label{sec:basis}

At least two iso-doublet scalar fields must be introduced in supersymmetric
theories to achieve a consistent formulation of the Higgs sector.
Supersymmetric theories are specific realizations of general scenarios which
include two doublets in the Higgs sector.  After three fields are absorbed to
generate the masses of the electroweak gauge bosons, five fields are left that
give rise to physical particles. In CP-invariant theories, besides the charged
states, two of the three neutral states are CP-even, while the third is
CP-odd. In CP-noninvariant theories the three neutral states however mix to
form a triplet with even and odd components in the wave--functions under CP
transformations \cite{kalinowski_gunion}-\cite{mssm_higgs_cp_violation}.  As
expected from general quantum mechanical rules, the mixing can become very
large if the states are nearly mass-degenerate. This situation is naturally
realized in supersymmetric theories in the decoupling limit in which two of
the neutral states are heavy.\s

In this note we analyze $H/A$ mixing in a simple quantum mechanical formalism
that reveals the underlying structures in a clear and transparent way.  $H$
and $A$ represent two heavy nearly mass--degenerate fields.  After the
discussion of the general CP--noninvariant 2--Higgs doublet model, we adopt
the Minimal Supersymmetric Standard Model, though extended to a
CP-noninvariant version [MSSM--CP], as a well--motivated example for the
analysis. \s\s

\section{Complex Mass Matrix}
\label{sec:complex_mass_matrix}

The most general form of the self-interaction of two Higgs doublets in a
CP--noninvariant theory is described by the potential
\cite{cp_invariant_decoupling}
\begin{eqnarray}
{\cal V}&=& m_{11}^2\Phi_1^\dagger\Phi_1+m_{22}^2\Phi_2^\dagger\Phi_2
           -[m_{12}^2\Phi_1^\dagger\Phi_2+{\rm h.c.}]\nonumber\\
        && +\frac{1}{2}\lambda_1(\Phi_1^\dagger\Phi_1)^2
           +\frac{1}{2}\lambda_2(\Phi_2^\dagger\Phi_2)^2
           +\lambda_3(\Phi_1^\dagger\Phi_1)(\Phi_2^\dagger\Phi_2)
           +\lambda_4(\Phi_1^\dagger\Phi_2)(\Phi_2^\dagger\Phi_1)\nonumber\\
        && +\left\{\frac{1}{2}\lambda_5(\Phi_1^\dagger\Phi_2)^2
           +\big[\lambda_6(\Phi_1^\dagger\Phi_1)
                +\lambda_7(\Phi_2^\dagger\Phi_2)\big]\,
                \Phi_1^\dagger\Phi_2+{\rm h.c.}\right\}
\label{eq:higgs_potential}
\end{eqnarray}
where $\Phi_{1,2}$ denote two complex $Y=1$, SU(2)$_L$ iso-doublet scalar
fields.  The coefficients are in general all non--zero. The parameters
$m^2_{12}, \lambda_{5,6,7}$ can be complex, incorporating the CP-noninvariant
elements in the interactions:
\begin{eqnarray}
m_{12}^2=m_{12}^{2R}+im_{12}^{2I}, \quad
\lambda_{5,6,7}=\lambda_{5,6,7}^R+i\lambda_{5,6,7}^I
\end{eqnarray}
Assuming the scalar fields to develop non-zero vacuum expectation values to
break the electroweak symmetries but leaving U(1)$_{EM}$ invariant, the vacuum
fields can be defined as
\begin{eqnarray}
\langle\Phi_1\rangle = \frac{v_1}{\sqrt{2}}
  \left(\begin{array}{c} 0 \\
                         1
        \end{array}\right), \quad
\langle\Phi_2\rangle = \frac{v_2}{\sqrt{2}}
  \left(\begin{array}{c} 0 \\
                         1
        \end{array}\right)
\end{eqnarray}
Without loss of generality, the two vacuum expectation values $v_i$ [$i=1,2$]
can be chosen real and positive after an appropriate global 
U(1) phase rotation; the
parameters of the (effective) potential 
Eq.(\ref{eq:higgs_potential}) are defined after this rotation.
As usual,
\begin{equation}
v=\sqrt{v_1^2+v_2^2} = 1/\sqrt{\sqrt{2} G_F} \;\; {\rm and} \;\;
\tan\beta=v_2/v_1
\end{equation}
with $v\approx 246$ GeV.
Abbreviations $t_{\beta}=\tan\beta$, $c_\beta=\cos\beta$,
$s_{2\beta} = \sin 2\beta$ etc, will be used from now on.\s

The conditions for minimizing the potential (\ref{eq:higgs_potential}) relate
the parameters $m^2_{ii}$ to the real part of $m^2_{12}$, $\lambda_k$, $v$
and $t_{\beta}$:
\begin{eqnarray}
&& m^2_{11} = m^{2R}_{12} t_\beta \,\,-\, \frac{1}{2} v^2 \left[\lambda_1
            c^2_\beta 
            + \lambda_{345} s^2_\beta + 3 \lambda^R_6 s_\beta c_\beta
            + \lambda^R_7 s^2_\beta t_\beta\right] \nonumber\\
&& m^2_{22} = m^{2R}_{12} t^{-1}_\beta-\frac{1}{2} v^2\left[\lambda_1 s^2_\beta
            + \lambda_{345} c^2_\beta + \lambda^R_6 c^2_\beta t^{-1}_\beta
            + 3\lambda^R_7 s_\beta c_\beta\right]
\end{eqnarray}
with the abbreviation $\lambda_{345} = \lambda_3 + \lambda_4 + \lambda^R_5$,
and the imaginary part of $m^2_{12}$ to the imaginary parts of the
$\lambda_{5,6,7}$ parameters:
\begin{eqnarray}
m^{2I}_{12} = \frac{1}{2} v^2 \left[\lambda^I_5 s_\beta c_\beta
             + \lambda^I_6 c^2_\beta + \lambda^I_7 s^2_\beta\right]
\end{eqnarray}
It will prove convenient later to exchange the real part of
$m^2_{12}$ for the auxiliary parameter $M^2_A$, or in units of $v$,
$m^2_A=M^2_A/v^2$, defined by the relation
\begin{eqnarray}
m^{2R}_{12} = \frac{1}{2}v^2 \left[ m^2_A s_{2\beta}
            + \lambda^R_5 s_{2\beta} + \lambda^R_6 c^2_\beta
                  +\lambda^R_7 s^2_\beta  \right]
\label{eq:ma}
\end{eqnarray}
This parameter will turn out to be one of the key parameters in the system.\s

In a first step the $\Phi_{1,2}$ system is rotated to the Higgs basis
$\Phi_{a,b}$,
\begin{eqnarray}
&& \Phi_a =\ \ \,  \cos\beta\, \Phi_1  + \sin\beta\, \Phi_2 \nonumber\\
&& \Phi_b = -\sin\beta\, \Phi_1  + \cos\beta\, \Phi_2
\end{eqnarray}
which is built up by the two iso--spinors:
\begin{eqnarray}
\Phi_a = \left(\begin{array}{c}
                G^+ \\
             \frac{1}{\sqrt{2}}
             \left(v+H_a+i G^0\right)
               \end{array}\right),
 \quad
\Phi_b= \left(\begin{array}{c}
                H^+ \\
             \frac{1}{\sqrt{2}}
             \left(H_b+i A\right)
              \end{array}\right)
\label{eq:ab_basis}
\end{eqnarray}
The three fields $G^{\pm,0}$ can be identified as the would-be Goldstone
bosons, while $H^{\pm}, H_{a,b}$ and $A$ give rise to physical Higgs bosons.
The charged Higgs mass $M_{H^\pm}$ and the real mass matrix ${\cal M}^2_{0R}$
of neutral Higgs fields in the basis of $H_a, H_b, A$ can easily be derived
from the potential after the rotations
\begin{eqnarray}
M^2_{H^{\pm}} = M^2_A+\frac{1}{2}\, v^2\lambda_F
\label{eq:charged-mass}
\end{eqnarray}
and
\begin{eqnarray}
{\cal M}^2_{0R}=v^2 \left(\begin{array}{ccc}
        \lambda     & -\hat{\lambda}            & -\hat{\lambda}_p \\[0.5mm]
        {   }       & \lambda-\lambda_A +m^2_A  & -\lambda_p \\[0.5mm]
        {   }       &     {   }                 &  m^2_A
                   \end{array}\right)
\label{eq:mass_matrix_ab}
\end{eqnarray}
to be abbreviated for easier reading and complemented symmetrically. The
notation for the real parts of the couplings,
\begin{eqnarray}
\lambda       &=& \lambda_1 c_{\beta}^4+\lambda_2 s_{\beta}^4
                 +\frac{1}{2}\lambda_{345}s_{2\beta}^2
                 +2s_{2\beta}(\lambda^R_6 c_{\beta}^2+\lambda^R_7 s_{\beta}^2)
               \nonumber\\
\hat{\lambda} &=& \frac{1}{2}s_{2\beta}\left[\lambda_1c_{\beta}^2
                 -\lambda_2s_{\beta}^2-\lambda_{345}c_{2\beta}\right]
                 -\lambda^R_6c_{\beta}c_{3\beta}-\lambda^R_7s_{\beta}s_{3\beta}
               \nonumber\\
\lambda_{A}   &=&  c_{2\beta}(\lambda_1c_{\beta}^2-\lambda_2s_{\beta}^2)
               +\lambda_{345}s_{2\beta}^2-\lambda^R_5
               +2\lambda^R_6c_{\beta}s_{3\beta}-2\lambda^R_7s_{\beta}c_{3\beta}
               \nonumber\\
\lambda_F     &=&  \lambda^R_5-\lambda_4
\label{eq:lambdas_definition}
\end{eqnarray}
essentially follows Ref.~\cite{cp_invariant_decoupling},  and
\begin{eqnarray}
&& \lambda_p       = \frac{1}{2}\lambda_5^I c_{2\beta}
                    -\frac{1}{2} (\lambda_6^I-\lambda_7^I)
                    s_{2\beta}\nonumber\\ 
&& \hat{\lambda}_p = \frac{1}{2}\lambda_5^I s_{2\beta}
                                 +\lambda_6^I c_{\beta}^2
                                 +\lambda_7^I s_{\beta}^2\,\,\,
                   [ = 2m_{12}^{2I}/ v^2 ]
\end{eqnarray}
are introduced for the imaginary parts of the couplings
\cite{cp_noninvariant}.\s

In a CP--invariant theory all couplings are real and the off-diagonal elements
$\lambda_p, \hat{\lambda}_p$ vanish. In this case the neutral mass matrix
separates into the standard CP-even $2\times 2$ part and the standard
[stand--alone] CP-odd part.\footnote{The Goldstone bosons $G^{\pm, 0}$
  (carrying zero mass) decouple from the physical states.}  The parameter
$M_A$ is then identified as the mass of the CP-odd Higgs boson $A$.  The
$2\times 2$ submatrix of the $H_a$ and $H_b$ system can be diagonalized,
leading to the two CP-even neutral mass eigenstates $h,~H$; in terms of
$H_a,H_b$:
\begin{eqnarray}
H &=& \cos\gamma\, H_a - \sin\gamma\, H_b\nonumber\\
h &=& \sin\gamma\, H_a + \cos\gamma\, H_b
\end{eqnarray}
with $\gamma = \beta- \alpha$; the angle $\alpha$ is the conventional CP--even
neutral Higgs boson mixing angle in the $[\Phi_1,\Phi_2]$ basis of the
CP--invariant 2HDM. The diagonalization of the mass matrix leads to the
relation:
\begin{eqnarray}
\tan 2\gamma=\frac{2\hat{\lambda}}{\lambda_A-m^2_A}
\end{eqnarray}
with $\gamma \in \left[0,\, \pi\right]$.\s

However, also in the general CP--noninvariant case, the fields $h_a = h,H,A$
serve as a useful basis, giving rise to the general final form of the real part
of the neutral mass matrix ${\cal M}^2_R$,
\begin{eqnarray}
{\cal M}^2_R = v^2 \left(\begin{array}{ccc}
      \lambda+(m^2_A-\lambda_A)\, c_{\gamma}^2 c_{2\gamma}^{-1}
    & 0
    & -\lambda_p\, c_{\gamma}-\hat{\lambda}_p\, s_{\gamma} \\[3mm]
      { }
    & \lambda-(m^2_A-\lambda_A)\, s_{\gamma}^2 c_{2\gamma}^{-1}
    & \lambda_p\, s_{\gamma}-\hat{\lambda}_p\, c_{\gamma} \\[3mm]
        {    }
    &   {    }
    & m^2_A         \end{array}\right)
\label{eq:cp_noninvariant_mass_matrix}
\end{eqnarray}
which is hermitian and symmetric by CPT invariance.\s

This hermitian part (\ref{eq:cp_noninvariant_mass_matrix}) of the mass matrix
is supplemented by the anti-hermitian part $-i M \Gamma$ incorporating the
decay matrix. This matrix includes the widths of the states $h_a=h,H,A$ in the
diagonal elements as well as the transition elements within any combination of
pairs. All these elements $(M\Gamma)^{AB}_{ab}$ are built up by loops of the
fields $(AB)$ in the self-energy matrix $\langle h_a h_b\rangle$ of the Higgs
fields.\s

In general, the light Higgs boson, the fermions and electroweak gauge bosons,
and in supersymmetric theories, gauginos, higgsinos and scalar states may
contribute to the loops in the propagator matrix. In the decoupling limit
explored later, the couplings of the heavy Higgs bosons to gauge bosons and
their supersymmetric partners are suppressed. Assuming them to decouple, being
significantly heavier, for example, than the Higgs states, we will consider
only loops built up by the light Higgs boson and the top quark as
characteristic examples; loops from other (s)particles could be treated in the
same way of course.\s\s

\noindent  
{\underline{(i) Light scalar Higgs $h$ states:}}\\[2mm]
While the $Hhh$ coupling is CP conserving, the $Ahh$ coupling is
CP--violating. Expressed in terms of the $\lambda$ parameters in the
potential they are given  as
\begin{eqnarray}
&& g_{Hhh}/v
    = -3\left( c_\beta c_\alpha s^2_\alpha \lambda_1
             + s_\beta s_\alpha c^2_\alpha \lambda_2\right)
      -\lambda_{345} \left[ c_\beta c_\alpha (3 c^2_\alpha-2)
                          + s_\beta s_\alpha (3 s^2_\alpha-2)\right]
       \nonumber\\[1mm]
&& { }\hskip 1.2cm
      -3 \lambda^R_6 \left[ s_\beta c_\alpha s^2_\alpha
                          + c_\beta s_\alpha (3 s^2_\alpha -2)\right]
n      -3 \lambda^R_7 \left[ c_\beta s_\alpha c^2_\alpha
                          + s_\beta c_\alpha (3 c^2_\alpha -2)\right]
     \nonumber\\[1mm]
&& g_{Ahh} /v=  \lambda^I_5 \left(s_\beta c_\beta - 2 s_\alpha c_\alpha \right)
          \nonumber\\[1mm]
&& { }\hskip 1.2cm
        + \lambda^I_6 \left[ (1 + 2 c_\beta^2) s_\alpha^2 - s_{2\beta}
                                                 s_\alpha c_\alpha \right]
        + \lambda^I_7 \left[ (1 + 2 s_\beta^2) c_\alpha^2 - s_{2\beta}
                                                 s_\alpha c_\alpha \right]
\end{eqnarray}
The trigonometric functions $s_\alpha$ and $c_\alpha$ can be re--expressed
by the sine and cosine of $\beta \;{\rm and}\; \gamma$
after inserting the difference $\alpha=\beta-\gamma$.
\s

The imaginary part of the light Higgs loop is given for CP--conserving and
CP--violating transitions by
\begin{eqnarray}
\left(M\Gamma\right)^{hh}_{HH/AA}
  &=& \frac{\beta_h}{32\pi} g^2_{Hhh/Ahh} \nonumber\\
\left(M\Gamma\right)^{hh}_{HA/AH}
  &=& \frac{ \beta_h}{32\pi} g_{Hhh} g_{Ahh}
\end{eqnarray}
where $\beta_h$ denotes the velocity of the light Higgs boson $h$ in the decays
$H/A \rightarrow hh$ [with the heavy Higgs bosons assumed to be
mass--degenerate].\s\s

\noindent
{\underline{(ii) Top--quark states:}}\\[2mm]
The $Htt$ and $Att$ couplings are defined by the Lagrangian
\begin{eqnarray}
{\cal L}_t = H \bar{t}\left[s_H + i\gamma_5 p_H\right] t
           + A \bar{t}\left[s_A+i\gamma_5 p_A\right] t
\label{eq:Htt_Att_couplings}
\end{eqnarray}
which includes the CP--conserving couplings $s_H, p_A$ and the CP--violating
couplings $p_H, s_A$. For the top quark loop we find
\begin{eqnarray}
\left(M\Gamma\right)^{tt}_{HH/AA}
   &=& \frac{3 M^2_{H/A}}{8\pi}\, \beta_t\, g^{tt}_{HH/AA} \nonumber\\
\left(M\Gamma\right)^{tt}_{HA/AH}
   &=& \frac{3 M^2_{H/A}}{8\pi}\, \beta_t\, g^{tt}_{HA/AH}
\end{eqnarray}
in the same notation as before. The transitions include incoherent and
coherent mixtures of scalar and pseudoscalar couplings,
\begin{eqnarray}
&& g^{tt}_{HH} = \beta^2_t s^2_H + p^2_H\nonumber\\[1mm]
&& g^{tt}_{AA} = \beta^2_t s^2_A + p^2_A\nonumber\\[1mm]
&& g^{tt}_{HA} = g^{tt}_{AH} = \beta^2_t s_H s_A + p_H p_A
\end{eqnarray}
where $\beta_t$ denotes the velocity of the top quarks in the Higgs rest
frame.  \s

These loops also contribute to the real part of the mass matrix. They either
renormalize the $\lambda$ parameters of the Higgs potential or they generate
such parameters if not present yet at the tree level. In the first case they
do not modify the generic form of the mass matrix, and the set of renormalized
$\lambda$'s are interpreted as free parameters to be determined
experimentally. The same procedure also applies to supersymmetric theories in
which some of the $\lambda$'s are generated radiatively by stop loops,
introducing CP--violation into the Higgs sector through bi-- and
trilinear--interactions in the superpotential, a case discussed later in
detail.\s

Including these elements, the final complex mass matrix can be written in the
Weisskopf-Wigner form \cite{ww_form}
\begin{eqnarray}
{\mathcal M}^2 = {\mathcal M}^2_R - i M \Gamma
\end{eqnarray}
which will be diagonalized in the next section. \s\s
\vskip 0.5cm

\noindent \underline{\bf Decoupling limit}. The decoupling limit
\cite{cp_invariant_decoupling} is defined by the inequality
\begin{eqnarray}
M^2_A\,\, \gg\,\, |\lambda_i|\, v^2\,
\end{eqnarray}
with $|\lambda_i| \lsim O(1)$ or $O(g^2, g^{\prime 2})$, $g^2$ and $g^{\prime
  2}$ denoting the electroweak gauge couplings. The limit is realized in many
supersymmetric models, particularly in SUGRA models with $M^2_A\gg v^2$.  It
is well known that in the decoupling limit the heavy states $H$ and $A$, as
well as $H^\pm$, are nearly mass degenerate. This feature is crucial for large
mixing effects between the CP--odd and CP--even Higgs bosons, $A$ and $H$,
analyzed in this report.\s

As the trigonometric $\sin$/$\cos$ functions of $\gamma=\beta-\alpha$ approach
the following values in the decoupling limit:
\begin{eqnarray}
c_\gamma\, \simeq\, \hat{\lambda}/m^2_A\, \rightarrow\, 0,\qquad
s_\gamma\, \rightarrow\, 1
\end{eqnarray}
up to second order in $1/m^2_A$, the real part of the complex
mass matrix acquires the simple form
\begin{eqnarray}
{\mathcal M}^2_R \,\simeq\, v^2\left(\begin{array}{ccc}
 \lambda   &         0                  &  -\hat{\lambda}_p \\[1.5mm]
 \cline{2-3}
  { }      & \multicolumn{1}{|c}{ } & { } \\[-4.mm]
  0        & \multicolumn{1}{|c}{m^2_A + \lambda-\lambda_A}
           & \lambda_p \\[1.5mm]
 -\hat{\lambda}_p &  \multicolumn{1}{|c}{\lambda_p}   &  m^2_A
                              \end{array}\right)
\label{eq:mass_matrix_decoupling}
\end{eqnarray}
in the leading order $\sim m^2_A$ and next--to--leading order $\sim 1$.
The $Hhh$ and $Ahh$ couplings are simplified in the
decoupling limit and they can be written in the condensed
form:
\begin{eqnarray}
&& \hskip -0.9cm g_{Hhh}/v \rightarrow -\frac{3}{2}
               s_{2\beta} \left(c^2_\beta \lambda_1
               - s^2_\beta \lambda_2-c_{2\beta} \lambda_{345}\right)
               \!+\! 3 \left(c_\beta c_{3\beta} \lambda^R_6
               + s_\beta s_{3\beta} \lambda^R_7\right)
               \,\rightarrow\, -3 \lambda^R_7  \nonumber\\
&& \hskip -0.9cm g_{Ahh}/v \rightarrow \frac{3}{2}
               s_{2\beta} \lambda^I_5
               \!+\! 3\left(c^2_\beta \lambda^I_6 + s^2_\beta
               \lambda^I_7\right) 
               \hskip 5.5cm
               \, \rightarrow\, +3 \lambda^I_7
\end{eqnarray}
In this limit we can set $c_\alpha= s_\beta$ and $s_\alpha=-c_\beta$. The
couplings simplify further for moderately large $\tan\beta$ and they are
determined in this range by the coefficient $\lambda_7$ alone as
demonstrated by the second column.\s

\section{Physical Masses and States}
\label{sec:decoupling_limit}

Following the steps in the appendix of Ref.\cite{nmssm_cmz}, it is easy to
prove mathematically, that mixing between the light Higgs state and the heavy
Higgs states is small, but large between the two nearly mass--degenerate
states.  Mathematically the mixing effects are of the order of the
off--diagonal elements in the mass matrix normalized to the difference of the
(complex) mass--squared eigenvalues. On quite general grounds, this is a
straightforward consequence of the uncertainty principle. We can therefore
restrict ourselves to the discussion of the mass degenerate $2 \times 2$
system of the heavy Higgs bosons $H,A$, allowing us to reduce the
calculational effort to a minimum.\s

If the mass differences become small, the mixing of the states is strongly
affected by the widths of the states and the complex Weisskopf--Wigner mass
matrix ${\cal M}^2={\cal M}^2_R-iM\Gamma$ must be considered in total, not
only the real part. This is well known in the literature from resonance mixing
\cite{resonant_mixing_zerwas} and has recently also been recognized for the
Higgs sector \cite{resonant_cp_pilaftsis}.\s

Since, by CPT invariance, the complex mass matrix ${\cal M}^2$ is symmetric,
adopting the notation in Ref.~\cite{resonant_mixing_zerwas} for the $H/A$
submatrix, separated in the lower right corner of
Eq.(\ref{eq:mass_matrix_decoupling}),
\begin{eqnarray}
{\mathcal M}^2_{HA}
  = \left(\begin{array}{cc}
   M^2_H - i M_H \Gamma_H  & \Delta^2_{HA}  \\[1mm]
    \Delta^2_{HA}        & M^2_A - i M_A \Gamma_A
    \end{array}\right)
\label{eq:complex_mass_matrix_squared}
\end{eqnarray}
the matrix can be diagonalized,
\begin{eqnarray}
{\mathcal M}^2_{H_i H_j}
  = \left(\begin{array}{cc}
   M^2_{H_2} - i M_{H_2} \Gamma_{H_2}  & 0                  \\[1mm]
      0      & M^2_{H_3} - i M_{H_3} \Gamma_{H_3}
    \end{array}\right)
\label{eq:complex_massij_matrix_squared}
\end{eqnarray}
through a {\it rotation by a complex mixing angle}:
\begin{eqnarray}
{\mathcal M}^2_{H_iH_j} = C {\mathcal M}^2_{HA} C^{-1},\quad
C = \left(\begin{array}{cc}
      \cos\theta  & \sin\theta \\[1mm]
     -\sin\theta  & \cos\theta
    \end{array}\right)
\end{eqnarray}
with
\begin{eqnarray}
X = \frac{1}{2}\tan 2\theta
  = \frac{\Delta^2_{HA}}{M^2_H - M^2_A
                          -i\left[M_H \Gamma_H - M_A\Gamma_A\right]}
\end{eqnarray}
A non--vanishing (complex) mixing parameter $\Delta^2_{HA}\neq 0$ requires
CP--violating transitions between $H$ and $A$ either in the real mass
matrix, $\lambda_p \neq 0$,
or in the decay mass matrix, $\Gamma_{HA} \neq 0$,
[or both]. When, in the decoupling limit,
the masses $M_H$ and $M_A$ are nearly degenerate, the widths may be
significantly different. Though nearly equal for decays to top pairs,
only the $H$ channel may be open for decays to light Higgs boson pairs.
As a result, the mixing phenomena are strongly affected
by the form of the decay matrix $M\Gamma$. This applies to the modulus
as well as the phase of the mixing parameter $X=\frac{1}{2}\tan 2\theta$.\s

The mixing shifts the Higgs masses and widths in a characteristic
pattern \cite{resonant_mixing_zerwas}. The two complex mass values
after and before diagonalization are
related by the complex mixing angle $\theta$:
\begin{eqnarray}
&& \hskip -0.7cm \left[M^2_{H_3}\!-\!i M_{H_3} \Gamma_{H_3}\right]
  \!\mp\!\left[M^2_{H_2}\!-\!i M_{H_2} \Gamma_{H_2}\right]
 \!=\!\left\{\left[M^2_A\!-\!i M_A \Gamma_A\right]
         \!\mp \!\left[M^2_H\!-\!i M_H \Gamma_H\right]\right\}
           \left\{ \hskip -0.2cm \begin{array} {l}
           { }\\[-6mm]
           \times \sqrt{1+4X^2} \\[6mm]
           \times \hskip 0.3cm 1
           \end{array}\right.  \nonumber\\[-0.61cm]
&&
\end{eqnarray}
As expected from rotation transformations, which leave the matrix spur
invariant, the complex eigenvalues split in exactly opposite directions when
the mixing is switched on.\footnote{At the very end of the analysis one may
  order the Higgs states according to ascending masses in CP--noninvariant
  theories. However, at intermediate steps the notation used here proves more
  transparent.}\s

The individual shifts of masses and widths can easily be obtained by separating
real and imaginary parts in the relations:
\begin{eqnarray}
 && \left[M^2_{H_2}\!-\!i M_{H_2} \Gamma_{H_2}\right]
  \!-\!\left[M^2_H\!-\!i M_H \Gamma_H\right]
   =
  - \left\{\left[M^2_{H_3}\!-\!i M_{H_3} \Gamma_{H_3}\right]
  \!-\!\left[M^2_A\!-\!i M_A \Gamma_A\right]\right\}\nonumber\\[2mm]
 &&{ }\hskip 2.32cm = -\left\{\left[M^2_A\!-\!i M_A \Gamma_A\right]
         \!-\!\left[M^2_H\!-\!i M_H \Gamma_H\right]\right\}
     \times\,\mbox{\small $\frac{1}{2}$}\, [\sqrt{1+4X^2}-1]
\end{eqnarray}
If the mixing parameter is small and real, the gap between the states increases
with the size of the mixing. \s

The eigenstates of the complex, non--hermitian matrix ${\cal M}^2_{HA}$ of
Eq.(\ref{eq:complex_mass_matrix_squared}) are no longer orthogonal, but
instead:
\begin{eqnarray}
\begin{array}{ll}
  |H_2\rangle =\ \ \, \cos\theta\, |H\rangle + \sin\theta\,\, |A\rangle, \ \
& \langle\, \widetilde{\!H}_2| =\ \ \, \cos\theta\, \langle H|
                            +\sin\theta\, \langle A| \\[1.5mm]
  |H_3\rangle = -\sin\theta\, |H\rangle + \cos\theta\,\, |A\rangle,\ \
& \langle\, \widetilde{\!H}_3| = -\sin\theta\, \langle H|
                            +\cos\theta\, \langle A|
\end{array}
\end{eqnarray}
Correspondingly, the final state $F$ in heavy Higgs formation from the initial
state $I$ is generated with the transition amplitude\footnote{Small
  off-resonant transitions of heavy Higgs bosons $H$ and $A$ to the light one
  $h$ in the decoupling limit (and to the neutral would-be Goldstone $G^0$)
  can be neglected to a good approximation.}
\begin{eqnarray}
\langle F|H|I\rangle = \sum_{i=2,3}\, \langle F| H_i \rangle\,
                       \frac{1}{s-M^2_{H_i}+i M_{H_i} \Gamma_{H_i}}\,
                       \langle\, \widetilde{\!H}_i|I\rangle
\end{eqnarray}

We illustrate the mixing mechanism in a simple toy model in which $M_A=0.5$
TeV, $\tan\beta=5$ and all $|\lambda_i|=0.4$ [{\it i.e.} roughly equal to the
weak SU(2) gauge coupling squared], while a common phase $\phi$ of all the
complex parameters $\lambda_{5,6,7}$ is varied from $0$ through $\pi$ to
$2\pi$.\footnote{With one common phase $\phi$, the complex mixing parameter
  $X$ obeys the relation $X(2\pi-\phi)=X^*(\phi)$, {\it i.e.} $\real/\imag
  X\rightarrow +\real/\!\!-\!\imag X$.  As a result, all CP--even quantities
  are symmetric and all CP--odd quantities anti--symmetric about $\pi$, {\it
    i.e.} when switching from $\phi$ to $2\pi-\phi$. Therefore we can restrict
  the discussion to the range $0\leq \phi\leq \pi$.}  The scalar and
pseudoscalar couplings of the top quark are identified with the standard
CP--conserving values $s_H\simeq p_A=\cot\beta \, m_t/v$ and $p_H=s_A=0$. The
mass of the light Higgs mass moves in this toy model from $M_h=215$ GeV to
$161$ GeV to $74$ GeV as the phase $\phi$ is varied from $0$ through $\pi/2$
to $\pi$ and, for $\phi=0$, the masses and widths of the heavy states are
$M_{H_2}=M_H= 520$ GeV, $M_{H_3} =M_A = 500$ GeV, $\Gamma_H=2.58$ GeV and
$\Gamma_A=1.49$ GeV. \s

\begin{figure}[tb!]
\begin{center}
\vskip 0.1cm
\includegraphics[width=13.0cm,height=13.5cm,clip=true]{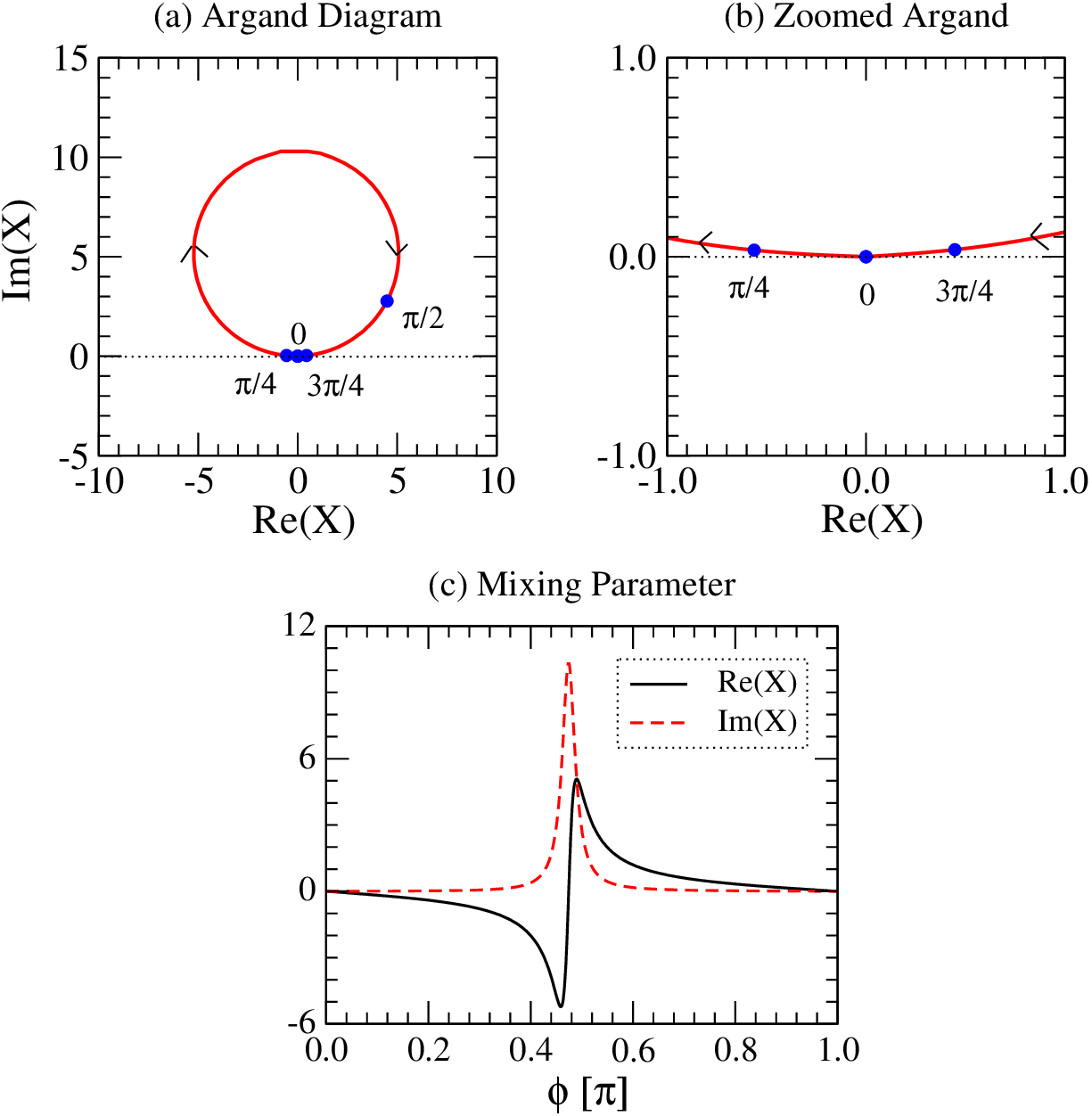}
\end{center}
\vskip -0.5cm
\caption{\it  (a,b) The Argand diagram and (c) the $\phi$ dependence of
     the mixing parameter $X$ in a toy model with the common
     CP--violating phase $\phi$ evolving from $0$ to $\pi$ for
     $\tan\beta=5$, $M_A=0.5$ TeV and with all $|\lambda_i|= 0.4$;
     the upper right--hand side zooms in on small angles.
     Note that $\real/\imag X(2\pi-\phi)=+\real/\!\!-\!\imag X(\phi)$. }
\label{fig:ReIm-toy}
\end{figure}

For these parameters, the Argand diagram of the mixing parameter $X$ is
presented in Fig.~\ref{fig:ReIm-toy}(a) in which the common CP--violating
phase $\phi$ evolves from $0$ to $\pi$ [for $\phi > \pi$ the reflection
symmetry $\real/\imag X \rightarrow +\real/\!\!-\!\imag X$ at $\phi = \pi$ may
be used]; Fig.~\ref{fig:ReIm-toy}(b) zooms in on the area of small angles.
Alternatively, the real and imaginary parts of $X$ are shown explicitly in
Fig.~\ref{fig:ReIm-toy}(c) as functions of the common CP--violating phase
$\phi$.\s

The difference of the squared masses $M^2_H-M^2_A$ and the real part of the
mass mixing parameter $\Delta^2_{HA}$ are approximately given by
\begin{eqnarray}
&& M^2_H - M^2_A = (\lambda - \lambda_A) v^2\,
     \approx\, \lambda v^2 \cos\phi
   \nonumber\\
&& \real(\Delta^2_{HA}) = \lambda_p v^2\,
\approx -\frac{1}{2}\lambda \, v^2 \sin\phi
\label{eq:matrel1}
\end{eqnarray}
and the imaginary parts by
\begin{eqnarray}
&& 32 \pi \left[M_H \Gamma_H - M_A \Gamma_A\right] \,\approx\, \Delta_t +
                                                 9\lambda^2 v^2 \cos{2\phi}
   \nonumber\\[1mm]
&& 32\pi\, \imag(\Delta^2_{HA})= 32 \pi M_{H/A} \Gamma_{HA}
               \,\approx\,- \frac{9}{2} \lambda^2 v^2 \sin{2\phi}
\label{eq:matrel2}
\end{eqnarray}
for the parameters specified above.  Since the complex couplings are
parameterized by a phase, $\cos\phi$ enters in the real part of the couplings
and thus affects the diagonal elements of the mass matrix. The difference of
the imaginary parts of the diagonal elements is determined by the widths of
the $H/A$ decays to top--quark pairs, $\Delta_t = - 12 M^2_{H/A} (m_t/v)^2 (1-
\beta_t^2) \beta_t$, modulated by sinusoidal variations from decays to $hh$.
The modulus of the real part of $X$ rises more rapidly than the imaginary
part; $|X|$ reaches unity for a phase $\sim \pi/3$, and the maximum value of
about $10$ a little below $\phi=\pi/2$ where $H$ and $A$ masses become equal.
The Argand diagram is described by a circle to a high degree of accuracy; the
center is located on the positive imaginary axis, and the radius of the circle
is given by $\sim \lambda v^2/4 |M_H \Gamma_H- M_A \Gamma_A| \sim 5$ in the
present scenario. Note that the resonant behavior is very sharp as shown in
Fig.\ref{fig:ReIm-toy}(c) which is also apparent from the swift move along the
circle in the Argand diagram.  The $\phi$ dependence of $X$ follows the
typical absorptive/dispersive pattern of analytical resonance amplitudes.\s

\begin{figure}[tb!]
\begin{center}
\vskip 0.2cm
\includegraphics[width=13.0cm,height=6.5cm,clip=true]{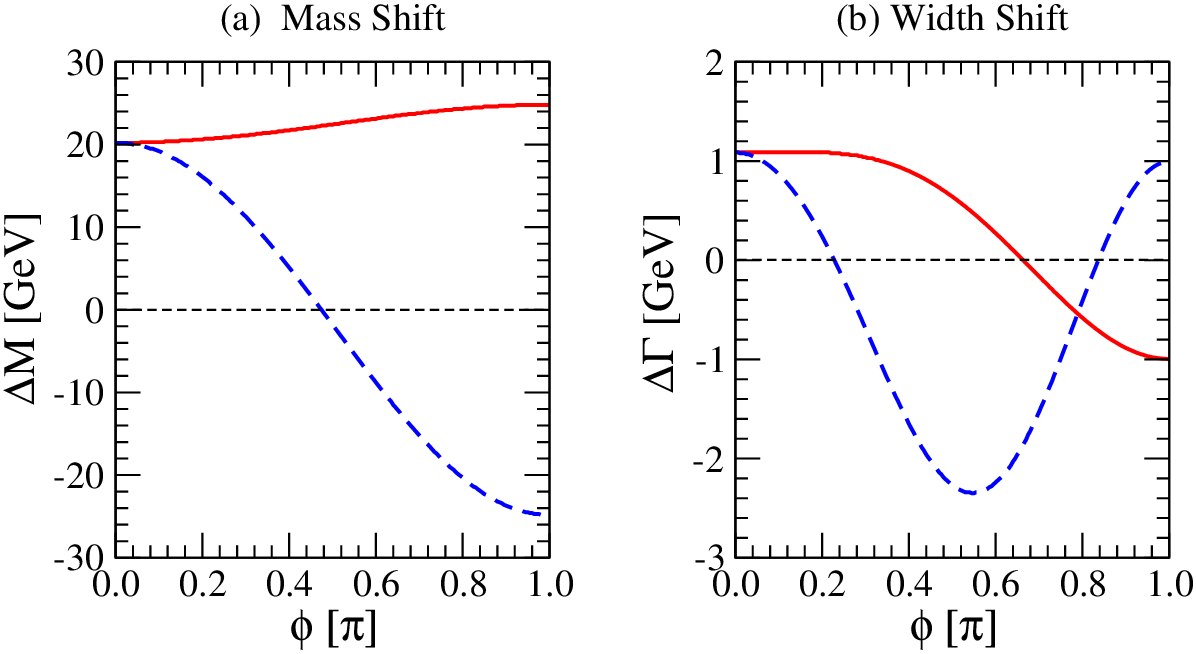}
\end{center}
\vskip -0.5cm
\caption{\it (a,b) The dependence of the mass and
     width shifts, $\Delta M=M_{H_2}- M_{H_3}$ and $\Delta\Gamma=\Gamma_{H_2}
     -\Gamma_{H_3}$, on the phase $\phi$. The dashed lines display these
     differences without mixing for the $H,A$ states. Both quantities are
     symmetric about $\phi=\pi$.}
\label{fig:shift}
\end{figure}

The shifts of masses and widths emerging from $H$ and $A$ are displayed in
Figs.~\ref{fig:shift}(a) and (b). The differences of masses and widths of $H$
and $A$ without the CP--violating mixing $\Delta^2_{HA}$ are shown by the
broken lines. As expected from Eq.(\ref{eq:matrel1}), the overall mass shift
decreases monotonically with varying $\phi$ from $0$ to $\pi$ while the width
shift shows an approximate sinusoidal behavior. If $\phi \approx \pi/2$ the
$H$--$A$ mass difference becomes so small that the mixing parameter $X$ can
become very large $\sim i \, \lambda v^2/ 2 \left(M_H \Gamma_H-M_A
  \Gamma_A\right) \sim 10\, i $ in the numerical example. Both CP--conserving
quantities are symmetric about $\phi=\pi$. The impact of $H/A$ mixing on the
character of $\Delta M$, in particular, is quite significant.\s

%

\section{A Specific SUSY Example}
\label{sec:specific_susy_example}

To illustrate these general quantum mechanical results in a potentially more
realistic example, we shall apply the formalism to a specific scenario within
the Minimal Supersymmetric Standard Model but extended by CP--violating
elements [MSSM--CP].  Following Ref.\cite{pilaftsis_wagner} we assume the SUSY
source of CP--violation to be localized in the superpotential with complex
higgsino parameter $\mu$ and trilinear coupling $A_t$ involving the top
squark.  All other interactions are assumed to be CP--conserving.\s

Through stop--loop corrections CP--violation is transmitted in this scenario
to the effective Higgs potential. Expressed in the general form
(\ref{eq:higgs_potential}), the effective $\lambda$ parameters have been
calculated in Ref.\cite{pilaftsis_wagner} to two--loop accuracy; to illustrate
the crucial points we recollect the compact one--loop results of the
$t/\tilde{t}$ contributions:
\begin{eqnarray}
&& \lambda_1 = \frac{g^2+g^{\prime 2}}{4}
           -\frac{h^4_t}{32\pi^2}\, \frac{|\mu|^4}{M^4_S}
           \qquad\hskip 2.7cm
   \lambda_5 = -\frac{h^4_t}{32\pi^2}\, \frac{\mu^2 A^2_t}{M^4_S}
           \nonumber\\[1mm]
&& \lambda_2 = \frac{g^2+g^{\prime 2}}{4}
           +\frac{3 h^4_t}{8\pi^2}\, \left[\log \frac{M^2_S}{m^2_t}
            + \frac{1}{2} X_t\right]
           \qquad\hskip 0.5cm
   \lambda_6 = \frac{h^4_t}{32\pi^2}\, \frac{|\mu|^2\mu A_t}{M^4_S}
           \nonumber\\[1mm]
&& \lambda_3 = \frac{g^2-g^{\prime 2}}{4}
           +\frac{h^4_t}{32\pi^2}\,
            \left(\frac{3|\mu|^2}{M^2_S}-\frac{|\mu|^2|A_t|^2}{M^4_S}\right)
           \quad
    \lambda_7 = -\frac{h^4_t}{32\pi^2}\,\frac{\mu}{M_S}\,
               \left(\frac{6A_t}{M_S}-\frac{|A_t|^2 A_t}{M^3_S}\right)
           \nonumber\\[1mm]
&& \lambda_4 = -\frac{g^2}{2}
           +\frac{h^4_t}{32\pi^2}\,
              \left(\frac{3|\mu|^2}{M^2_S}-\frac{|\mu|^2|A_t|^2}{M^4_S}\right)
\label{eq:lambda_mssm}
\end{eqnarray}
where
\begin{eqnarray}
h_t = \frac{\sqrt{2} \overline{m}_t (m_t)}{v \sin\beta}\qquad  {\rm and}\qquad
X_t = \frac{2|A_t|^2}{M^2_S} \left(1-\frac{|A_t|^2}{12 M^2_S}\right)
\end{eqnarray}
Here, $m_t$ is the top--quark pole mass related to the running $\overline{MS}$
mass $\overline{m}_t(m_t)$ through $\overline{m}_t (m_t) =
m_t/[1+\frac{4}{3\pi} \alpha_s(m_t)]$, and $M_S$ is the SUSY scale.\s

To demonstrate the complex $H/A$ mixing in this MSSM--CP model numerically, we
adopt a typical set of parameters from
Refs.\cite{mssm_higgs_cp_violation,cpsuperh},
\begin{eqnarray}
M_S = 0.5\,\, {\rm TeV},\quad |A_t|= 1.0\,\, {\rm TeV},\quad
|\mu|= 1.0\,\, {\rm TeV},\quad \phi_\mu=0\,; \quad \tan\beta=5
\end{eqnarray}
while varying the phase $\phi_A$ of the trilinear parameter
$A_t$.\footnote{Analyses of electric dipole moments strongly suggest that CP
  violation in the higgsino sector will be very small in the MSSM--CP
  \cite{edm}; therefore we set $\phi_\mu=0$. Note that the $\lambda$'s in
  Eq.(\ref{eq:lambda_mssm}) are actually affected only by one common phase
  which is the sum of the angles $(\phi_A+\phi_\mu)$.}  We find the following
mass values of the light and heavy Higgs masses in the CP--conserving case
with $\phi_A=0$:
\begin{eqnarray}
M_h = 129.6\,\, {\rm GeV}, \quad M_H = 500.3\,\, {\rm GeV},\quad
M_A = 500.0\,\, {\rm GeV}
\end{eqnarray}
and their widths:
\begin{eqnarray}
\Gamma_H = 1.2\,\, {\rm GeV},\qquad
\Gamma_A = 1.5\,\, {\rm GeV}
\end{eqnarray}
\begin{figure}[tb!]
\begin{center}
\vskip 0.2cm
\includegraphics[width=13.0cm,height=13.5cm,clip=true]{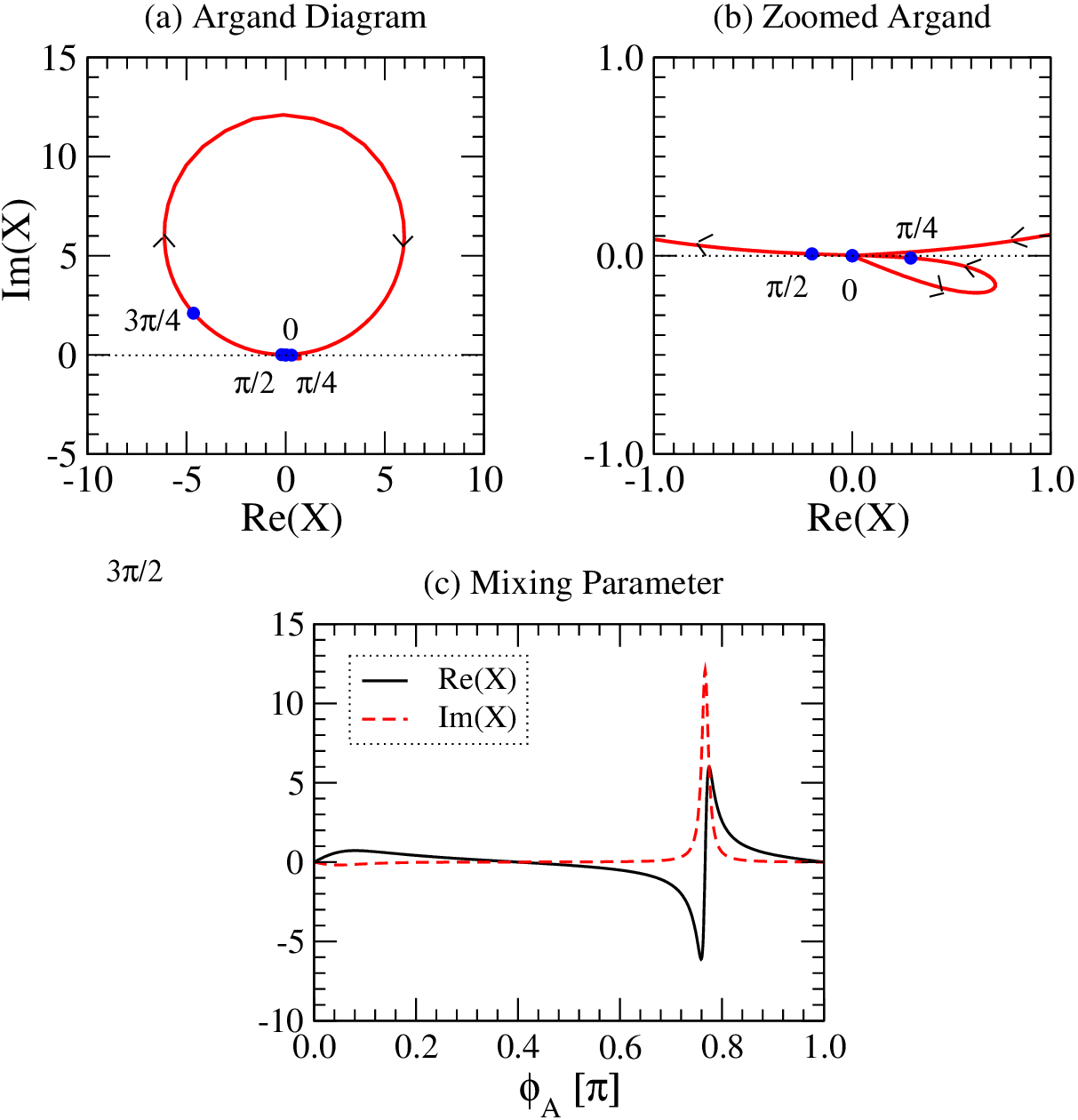}
\end{center}
\vskip -0.7cm
\caption{\it   (a,b) The Argand diagram  and (c) the $\phi_A$ dependence of
     the mixing parameter $X$ in a SUSY model with the
     CP--violating phase $\phi_A$ evolving from $0$ to $\pi$ for
     $\tan\beta=5$, $M_A=0.5$ TeV and couplings as specified in
     the text; the Argand diagram zoomed in on small angles is
     displayed on the upper right--hand frame. $\real/\imag X(2\pi-\phi_A)
     =+\real/\!\!-\!\imag X(\phi_A)$ for angles above $\pi$.}
\label{fig:ReIm-mssm}
\end{figure}

\begin{figure}[th!]
\begin{center}
\vskip 0.0cm
\includegraphics[width=13.2cm,height=6.5cm]{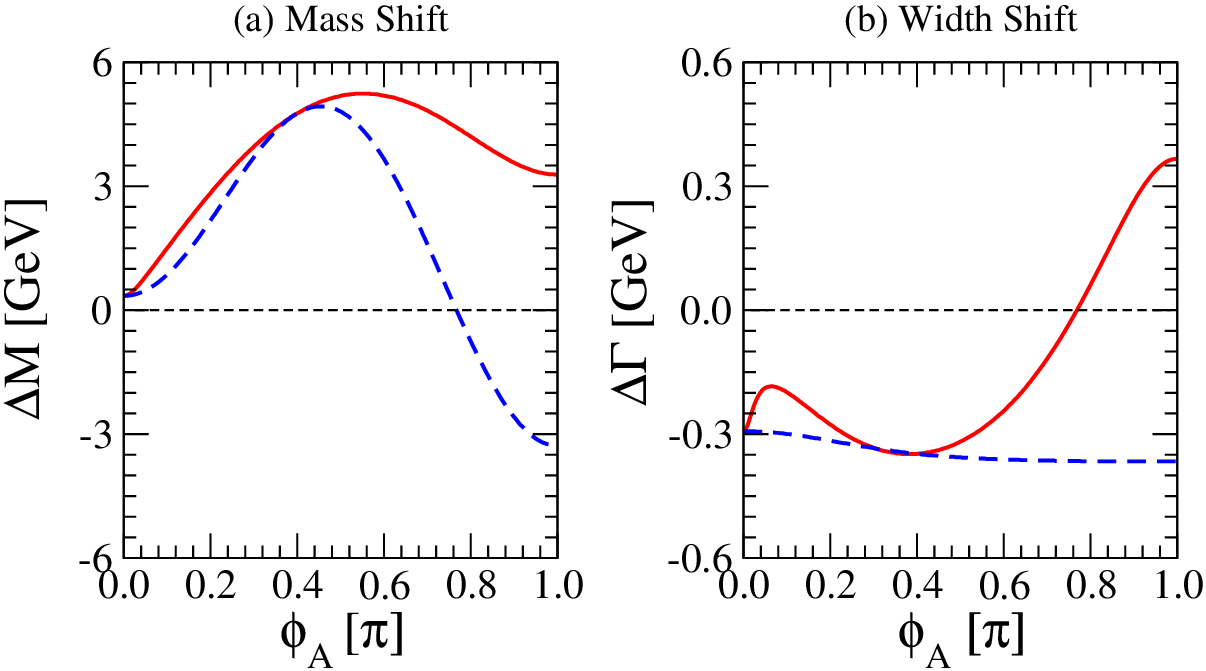}
\end{center}
\vskip -0.4cm
\caption{\it (a,b) The dependence of the shifts of masses and widths on
         the CP-violating angle $\phi_A$ in the SUSY model with the
         same parameter set as in Fig.\ref{fig:ReIm-mssm}; the differences
         without mixing are shown by the broken lines.}
\label{fig:shift_mssm}
\end{figure}

While the light Higgs boson mass is not altered if CP--violation through the
phase $\phi_A$ is turned on, the Argand diagram and the variation of the
CP--violating parameter $X$ are presented in Figs.~\ref{fig:ReIm-mssm}(a), (b)
and (c). [Symmetries in moving from $\phi_A$ to $2\pi-\phi_A$ are identical to
the toy model.]  The mass and width shifts of the heavy neutral Higgs bosons
are displayed in Figs.\ref{fig:shift_mssm}(a) and (b), respectively. Similar
to the toy model in the previous section, the two--state system in the
MSSM--CP shows a very sharp resonant CP--violating mixing, purely imaginary, a
little above $\phi_A = 3\pi/4$.  The mass shift is enhanced by more than an
order of magnitude if the CP--violating phase rises to non-zero values,
reaching a maximal value of $\sim 5.3$ GeV; the width shift moves up
[non-uniformly] from $-0.3$ and $+0.4$ GeV. As a result, the two
mass--eigenstates become clearly distinguishable, incorporating significant
admixtures of CP--even and CP--odd components mutually in the
wave--functions.\s

%

\section{Experimental Signatures of CP Mixing }
\label{eq:two_examples}

{\bf (i)} A first interesting example for studying CP--violating mixing
effects is provided by {\it \underline{$\gamma\gamma$--Higgs formation in
polarized beams}} \cite{general_method,gg_h,TESLA}:
\begin{eqnarray}
\gamma\gamma \rightarrow H_i \ \ [i=2,3]
\end{eqnarray}
For a specific final state $F$ of the Higgs boson decays, the amplitude of the
reaction $\gamma\gamma\rightarrow H_i\rightarrow F$ is a superposition of
$H_2$ and $H_3$ exchanges. For helicities $\lambda=\pm 1$ of the two photons,
the amplitude reads
\begin{equation}
{\cal M}^F_{\lambda}
  =\sum_{i=2,3}\, \langle F| H_i\rangle\, D_i(s) \,
  \left[ S^\gamma_i (s)+i\lambda P^\gamma_i(s) \right]
\end{equation}
The loop--induced $\gamma\gamma H_i$ amplitudes are described by the scalar
and pseudoscalar form factors, $S^\gamma_i(s)$ and $P^\gamma_i(s)$ where
$\sqrt{s}$ is the $\gamma\gamma$ energy and the Higgs $H_i$ propagator
$D_i(s)=1/(s-M^2_{H_i}+i M_{H_i}\Gamma_{H_i})$.  The scalar and pseudoscalar
form factors receive the {\it dominant} contributions from the top (s)quark
loops in the decoupling regime for moderate values of $\tan\beta$. They are
related to the well--known conventional $\gamma\gamma H/A$ form factors,
$S^\gamma_{H,A}$ and $P^\gamma_{H,A}$, by
\begin{eqnarray}
&& S^\gamma_2=\cos\theta\, S^\gamma_H + \sin\theta\, S^\gamma_A\qquad\,\,
   S^\gamma_3=-\sin\theta\, S^\gamma_H+ \cos\theta\, S^\gamma_A
   \nonumber\\[2mm]
&& P^\gamma_2=\cos\theta\, P^\gamma_H + \sin\theta\, P^\gamma_A\qquad
   P^\gamma_3=-\sin\theta\, P^\gamma_H+ \cos\theta\, P^\gamma_A
\label{eq:ggh_form_factor}
\end{eqnarray}
For the explicit form of the loop functions $S^\gamma_{H,A}$ and
$P^\gamma_{H,A}$ see, for example, Ref.\cite{cpsuperh}.
To reduce technicalities we will assume
from now on that the Higgs--$tt$ couplings are CP--conserving, {\it i.e.}
$P^\gamma_H$ and $S^\gamma_A = 0$.
The production rates of heavy SUSY Higgs bosons in such a scenario
have been calculated in Ref.\cite{muhl}.   \s

For linearly polarized photons, the CP--even component of the $H_i$
wave--functions is projected out if the polarization vectors are parallel,
and the CP--odd component if they are perpendicular.
This effect can be observed in the
CP--even asymmetry
\begin{eqnarray}
{\cal A}_{lin}
  = \frac{\sigma_\parallel- \sigma_\perp}{\sigma_\parallel+ \sigma_\perp}
\end{eqnarray}
of the total $\gamma\gamma$ fusion cross sections for linearly polarized
photons. Though not CP--violating {\it sui generis}, the asymmetry ${\cal
  A}_{lin}$ provides us with a powerful tool nevertheless to probe
CP--violating admixtures to the Higgs states since $|{\cal A}_{lin}|<1$
requires both $S^\gamma_i$ and $P^\gamma_i$ non-zero couplings. Moreover,
CP--violation due to $H/A$ mixing can directly be probed via the CP--odd
asymmetry\footnote{This asymmetry is also odd under CP$\tilde{\rm T}$ where
  the naive time reversal transformation $\tilde{\rm T}$ \cite{naive_T}
  reverses the direction of all 3--momenta and spins, but does not exchange
  initial and final state. Quantities that are odd under CP$\tilde{\rm T}$ can
  be non--zero only for complex transition amplitudes with absorptive
  phases which can be generated, for example, by loops, and Breit-Wigner
  propagators.}  constructed with circular photon polarization as
\begin{eqnarray}
{\cal A}_{hel}
  =\frac{\sigma_{++} -\sigma_{--}}{\sigma_{++}+ \sigma_{--}}
\end{eqnarray}

\begin{figure}[th!]
\begin{center}
\vskip -0.1cm
\includegraphics[width=14.0cm,height=13.cm,clip=true]{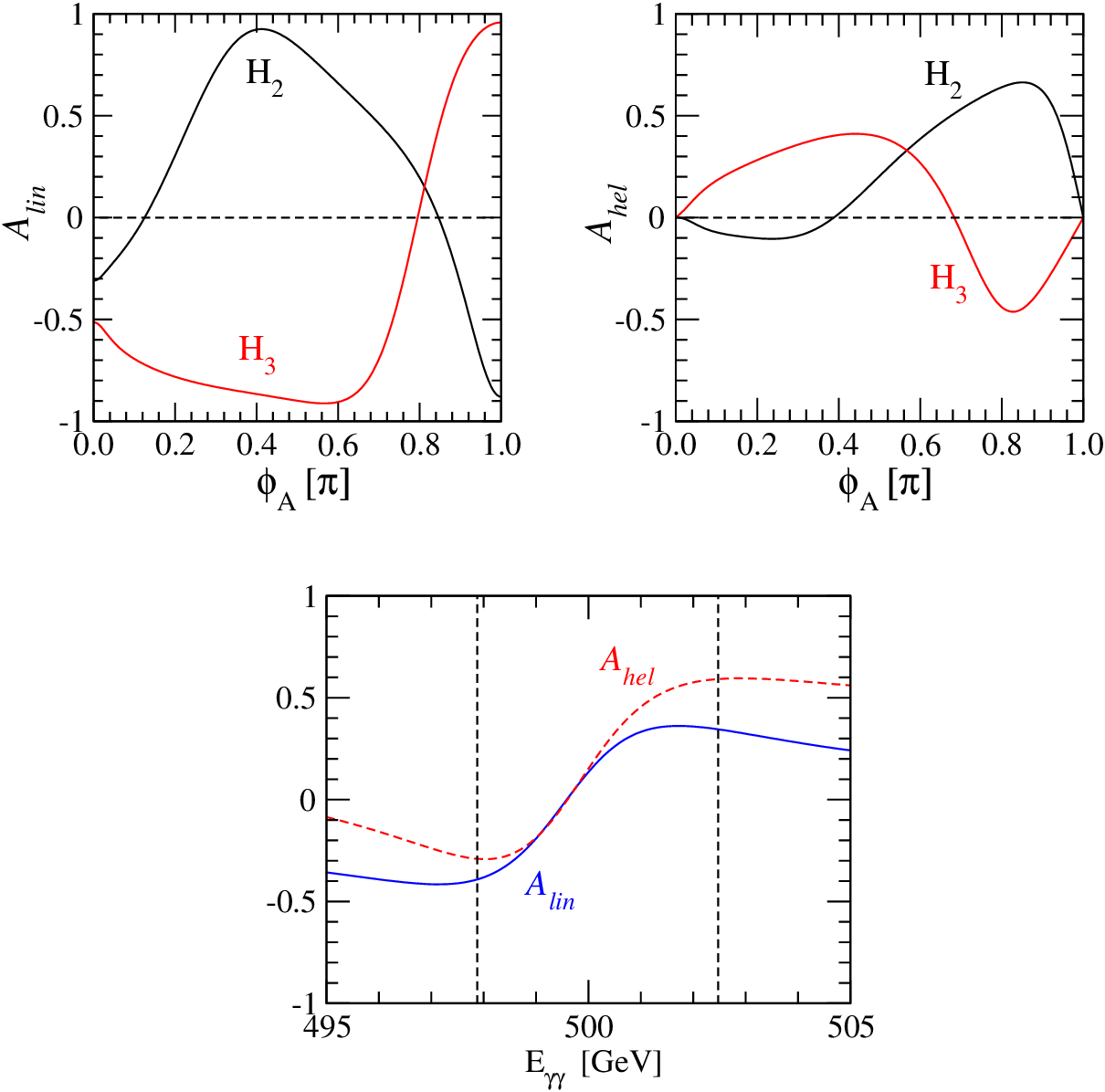}
\end{center}
\vskip -0.7cm
\caption{\it The $\phi_A$ dependence of the CP--even and CP--odd correlators,
         ${\cal A}_{lin}$ (upper--left panel) and ${\cal A}_{hel}$
         (upper--right 
         panel), at the pole of $H_2$ and $H_3$, respectively, and the
         $\gamma\gamma$ energy dependence (lower panel) of the correlators,
         ${\cal A}_{lin,hel}$ for $\phi_A=3\pi/4$ in the production process
         $\gamma\gamma \rightarrow H_i$. The same parameter set as in
         Fig.\ref{fig:ReIm-mssm} is employed. Numerically, $M_{H_2}=502.5$ GeV,
         $M_{H_3}=497.9$ GeV, $\Gamma_{H_2}=1.28$ GeV and $\Gamma_{H_3}=1.31$
         GeV. The vertical lines on the lower panel represent the two mass
         eigenvalues, $M_{H_3}$ and $M_{H_2}$.}
\label{fig:asym_rrh}
\end{figure}

The upper panels of Fig.\ref{fig:asym_rrh} show the $\phi_A$ dependence of the
asymmetries ${\cal A}_{lin}$ and ${\cal A}_{hel}$ at the pole of $H_2$ and of
$H_3$, respectively, for the same parameter set as in Fig.\ref{fig:ReIm-mssm}
and with the common SUSY scale $M_{\tilde{Q}_3}=M_{\tilde{t}_R}= M_S=0.5$ TeV
for the soft SUSY breaking top squark mass parameters.\footnote{On quite
  general grounds, the CP--conserving observables are symmetric under the
  reflection about $\phi_A=\pi$, while the CP--violating observables are
  anti--symmetric.}  By varying the $\gamma\gamma$ energy from below $M_{H_3}$
to above $M_{H_2}$, the asymmetries, ${\cal A}_{lin}$ (blue solid line) and
${\cal A}_{hel}$ (red dashed line), vary from $-0.39$ to $0.34$ and from
$-0.29$ to $0.59$, respectively, as demonstrated on the lower panel of
Fig.\ref{fig:asym_rrh} with $\phi_A=3\pi/4$, a phase value close to resonant
CP--mixing.\s

If the widths are neglected, the asymmetries ${\cal A}_{lin}$ and ${\cal
  A}_{hel}$ on top of the $H_i [i=2,3]$ resonances can approximately be
written in terms of the form factors:
\begin{eqnarray}
&& {\cal A}_{lin} [H_i]
     \,\approx \, \frac{|S_i^\gamma|^2-|P_i^\gamma|^2}
                  {|S_i^\gamma|^2+|P_i^\gamma|^2}\\[2mm]
&& {\cal A}_{hel} [H_i]
     \,\approx \, \frac{2\, {\imag} (S_i^\gamma P_i^{\gamma\ast})}
                  {|S_i^\gamma|^2+|P_i^\gamma|^2}
\end{eqnarray}
These approximate formulae reproduce the numerical values very accurately.  If
one further neglects not only small corrections due to such overlap phenomena
but also corrections due to non-asymptotic Higgs-mass values, the asymmetries
on top of the resonances $H_2$ and $H_3$ can simply be expressed by the
complex mixing angle $\theta$:
\newpage
\begin{eqnarray}
 {\cal A}_{lin} [H_2] \, \simeq \, - {\cal A}_{lin} [H_3]
     \, \simeq\, \frac{|\cos\theta|^2-|\sin\theta|^2}{
                  |\cos\theta|^2+|\sin\theta|^2}\\[1mm]
 {\cal A}_{hel} [H_2] \, \simeq \, + {\cal A}_{hel} [H_3]
     \, \simeq\, \frac{2\, \imag (\cos\theta \sin\theta^\ast)}
                             {|\cos\theta|^2+|\sin\theta|^2}
\end{eqnarray}
The $ {\cal A}_{lin}$ asymmetries are opposite in sign for the two Higgs
bosons $H_2$ and $H_3$, while the ${\cal A}_{hel}$ have the same sign.
However, we note that the corrections due to non--asymptotic Higgs masses are
still quite significant for the mass ratio $M_{H_2,H_3} / 2 m_t \sim 1.3$ in
our reference point, particularly for ${\cal{A}}_{hel}$ which is sensitive to
the interference between the ${\gamma\gamma}H$ and ${\gamma\gamma}A$ form
factors{\footnote{We have checked that indeed the numerical values approach
    formula (50) for very large Higgs masses.}.\s

Detailed experimental simulations would be needed to estimate the accuracy
with which the asymmetries can be measured. However, the large magnitude and
the rapid, significant variation of the CP--even and CP--odd asymmetries,
${\cal A}_{lin}$ and ${\cal A}_{hel}$, through the resonance region with
respect to both the phase $\phi_A$ and the $\gamma\gamma$ energy would be a
very interesting effect to observe in any case.\s\s \vskip 0.3cm

\noindent
{\bf (ii)} A second observable of interest for studying CP--violating mixing
effects experimentally is the {\it \underline{polarization of the top quarks
    in $H_i$ decays}} produced by $\gamma\gamma$ fusion
\cite{general_method,h_tt} or elsewhere:
\begin{eqnarray}
H_i \rightarrow t\bar{t}\quad \ [i=2,3]
\end{eqnarray}
Even if the $H/A tt$ couplings are CP--conserving, the complex rotation
matrix $C$ may mix the CP--even $H$ and CP--odd $A$ states leading to the
CP--violating helicity amplitude of the decay process $H_i\rightarrow
t\bar{t}$:
\begin{eqnarray}
\langle t_\sigma \bar{t}_\sigma | H_i\rangle = \sum_{a=H, A}
        C_{ia} (\sigma \beta_t s_a - i p_a)
\end{eqnarray}
where the $t$ and $\bar{t}$ helicities $\sigma/2=\pm 1/2$ must be
equal and $s_a, p_a$ are the $Htt$ and $Att$ couplings defined in
Eq.(\ref{eq:Htt_Att_couplings}). The two correlations between the transverse
$t$ and $\bar{t}$ polarization vectors $s_{\bot},\bar{s}_{\bot}$
in the production--decay process $\gamma\gamma\rightarrow H_i\rightarrow
t\bar{t}$,
\begin{eqnarray}
{\cal C}_\parallel = \left\langle s_\perp \cdot \bar{s}_\perp
                 \right\rangle\qquad {\rm and}\qquad
{\cal C}_\perp = \left\langle \hat{p}_t\cdot (s_\perp\times\bar{s}_\perp)
                 \right\rangle
\end{eqnarray}
lead to a non--trivial CP--even [CP$\tilde{\rm T}$--even] azimuthal
correlation and a CP--odd [CP$\tilde{\rm T}$--even] azimuthal correlation,
respectively, between the two decay planes $t\rightarrow bW^+$ and
$\bar{t}\rightarrow \bar{b}W^-$:
\begin{eqnarray}
\frac{1}{\Gamma}\, \frac{d\Gamma}{d\phi^*} \,
    = \,  \frac{1}{2\pi}\,
 \left[ 1 -\frac{\pi^2}{16}\left(
          \frac{1-2 m^2_W/m^2_t}{1+2 m^2_W/m^2_t}\right)^2\, \left(
          {\cal C}_\parallel\, \cos\phi^*
        + {\cal C}_\perp\, \sin\phi^*\right) \right]
\end{eqnarray}
where $\phi^*$ denotes the azimuthal angle between two decay planes
\cite{general_method}. In terms of the helicity amplitudes
$\langle\sigma,\lambda\rangle$ for the process $\gamma\gamma\rightarrow
H_i\rightarrow t\bar{t}$, where $\lambda=\pm 1$ denotes the helicities of both
photons and $\sigma=\pm 1$ twice the helicities of both top quarks, the
asymmetries are given as
\begin{eqnarray}
&&{\cal C}_\parallel = -\, \frac
   {2\,\real \sum \langle+,\lambda\rangle
                  \langle-,\lambda\rangle^*}
   {\sum \left(|\langle+,\lambda\rangle|^2
                      +|\langle-,\lambda\rangle|^2\right)}
   \\[2mm]
&&{\cal C}_\perp     =+\, \frac
   {2\, \imag\sum \langle+,\lambda\rangle\langle-,\lambda\rangle^*}
   {\sum \left(|\langle+,\lambda\rangle|^2
                      +|\langle-,\lambda\rangle|^2\right)}
\end{eqnarray}
with the sum running over the two photon helicities.\s

The upper panels of Fig.\ref{fig:asym_htt} show the $\phi_A$ dependence of the
CP--even and CP--odd asymmetries, ${\cal C}_\parallel$ and ${\cal C}_\perp$,
at the pole of $H_2$ and of $H_3$, respectively, for the same parameter set as
in Fig.\ref{fig:asym_rrh}. If the invariant $t\bar{t}$ energy is varied
throughout the resonance region, the correlators ${\cal C}_\parallel$ (blue
solid line) and ${\cal C}_\perp$ (red dashed line) vary characteristically
from $-0.43$ to $-0.27$ [non--uniformly] and from $0.84$ to $-0.94$,
respectively, as shown on the lower panel of Fig.\ref{fig:asym_htt}.\s

\begin{figure}[ht!]
\begin{center}
\vskip -0.1cm
\includegraphics[width=14.0cm,height=13.cm,clip=true]{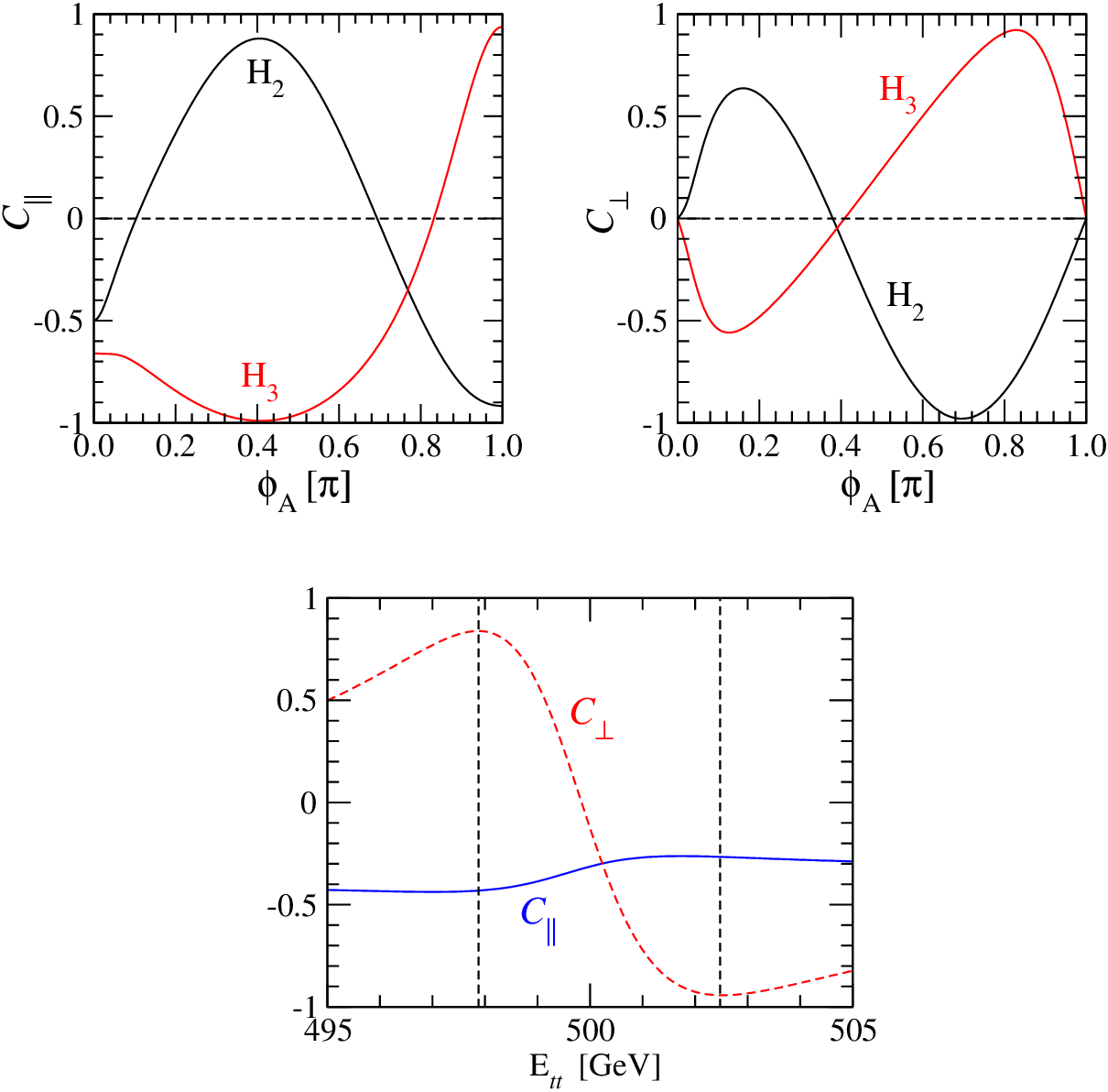}
\end{center}
\vskip -0.6cm
\caption{\it The $\phi_A$ dependence of the CP--even and CP--odd correlators,
         ${\cal C}_\parallel$ (upper--left panel)  and ${\cal C}_\perp$
         (upper--right panel), at the pole of $H_2$ and $H_3$ and the invariant
         $t\bar{t}$ energy dependence (lower panel) of the correlators
         ${\cal C}_{\parallel,\perp}$ for $\phi_A=3\pi/4$ in the
         production--decay chain $\gamma\gamma \rightarrow H_i\rightarrow
         t\bar{t}$. [Same SUSY parameter set as in Fig.\ref{fig:asym_rrh}.]}
\label{fig:asym_htt}
\end{figure}

Similarly to the previous case, if the widths are neglected,
the ${\cal C}_\parallel$ and ${\cal C}_\perp$ asymmetries on top of the
resonances $H_2$ and $H_3$ can approximately be expressed by the complex
mixing angle $\theta$ as:
\begin{eqnarray}
&& {\cal C}_\parallel [H_2]
    \,\simeq\,
  \frac{|\cos\theta|^2 \beta^2_t-|\sin\theta|^2}{
        |\cos\theta|^2 \beta^2_t+|\sin\theta|^2}\quad\ \
 {\cal C}_\parallel [H_3]
    \,\simeq\, -
  \frac{|\cos\theta|^2-|\sin\theta|^2 \beta^2_t}{
        |\cos\theta|^2+|\sin\theta|^2 \beta^2_t}\\[2mm]
&& {\cal C}_\perp [H_2]
    \,\simeq\,
  \frac{2 \real(\cos\theta \sin\theta^*) \beta_t}{
        |\cos\theta|^2 \beta^2_t+|\sin\theta|^2}\, \quad\,
 {\cal C}_\perp [H_3]
    \,\simeq\, -
  \frac{2 \real(\cos\theta \sin\theta^*) \beta_t}{
        |\cos\theta|^2+|\sin\theta|^2 \beta^2_t}
\end{eqnarray}
These approximate formulae provide a nice qualitative understanding of the
numerical values. In the asymptotic kinematic limit $\beta_t \to 1$ of the
top-quark speed, the correlators reduce to the simple expressions:
\begin{eqnarray}
&& {\cal C}_\parallel [H_2]
    \,\simeq\,
   -  {\cal C}_\parallel [H_3]
    \, \simeq \,
  \frac{|\cos\theta|^2-|\sin\theta|^2}{
        |\cos\theta|^2+|\sin\theta|^2}\qquad\quad\,\,\,\,\\[2mm]
&& {\cal C}_\perp [H_2]
    \,\simeq\,
    - {\cal C}_\perp [H_3]
    \, \simeq \,
    \frac{2\, \real(\cos\theta \sin\theta^*)}{
        |\cos\theta|^2+|\sin\theta|^2}
\end{eqnarray}
{\it i.e.} they are both opposite in sign. However, we note that the square of
the top--quark speed $\beta^2_t\approx 0.5$ near the Higgs resonances so that
the corrections due to non--asymptotic Higgs masses are significant, in
particular, for the asymmetry ${\cal C}_\parallel$ in the present example.\s

Though not easy to observe, the gross effects, at least, in
Fig.\ref{fig:asym_htt} should certainly be accessible experimentally. \s\s

\section{CONCLUSIONS}
\label{sec:conclusion}

Exciting mixing effects can occur in the Higgs sector of two--Higgs
doublet models, {\it notabene} in supersymmetric models, if CP--noninvariant
interactions are switched on. In the decoupling regime these effects can
become very large, leading to interesting experimental consequences. We have
analyzed such scenarios in quite a general quantum mechanical language that
provides us with a clear and transparent understanding of the phenomena
in the general 2--doublet model. Moreover, the effects are illustrated in the
Minimal Supersymmetric Standard Model extended by CP violating interactions
[MSSM--CP]. Higgs formation in $\gamma\gamma$ collisions proves particularly
interesting for observing such effects. However, exciting experimental effects
are also predicted in such scenarios  for $t \bar{t}$ final--state analyses
in decays of the heavy Higgs bosons at LHC and in the $e^+e^-$ mode of linear
colliders.

\vskip 0.5cm

\subsection*{Acknowledgements}
We are grateful to Grahame Blair for valuable comments on the
manuscript.
The work is supported in part by the European Commission 5-th Framework
Contract HPRN-CT-2000-00149.
SYC was supported in part by a Korea Research Foundation Grant
(KRF--2002--070--C00022) and in part by KOSEF through CHEP at Kyungpook
National University. JK was supported by the KBN Grant 2 P02B 040 20
(2003-2005) and 115/E-343/SPB/DESY/P-03/DWM517/2003-2005.

\vskip 1.cm

\end{document}